\documentclass[twocolumn]{aastex631}
\usepackage{amsmath}

\usepackage{enumitem}

\begin{document}

\title{A Possible Four-Month Periodicity in the Activity of FRB 20240209A}

\author[0009-0007-8409-4233]{Arpan Pal}
\affiliation{National Centre for Radio Astrophysics, Tata Institute of Fundamental Research, S. P. Pune University Campus, Ganeshkhind, Pune, 411007}

\begin{abstract}
Fast Radio Bursts (FRBs) are millisecond-duration radio transients from distant galaxies. While most FRBs are singular events, repeaters emit multiple bursts, with only two—FRB 121102 and FRB 180916B-showing periodic activity (160 and 16 days, respectively). FRB 20240209A, discovered by CHIME-FRB, is localized to the outskirts of a quiescent elliptical galaxy (z = 0.1384). We discovered a periodicity of $\sim 126$ days in the activity of the FRB 20240209A, potentially adding to the list of extremely rare periodic repeating FRBs. We used auto-correlation and Lomb-Scargle periodogram analyses, validated with randomized control samples, to confirm the periodicity. The FRB's location in an old stellar population disfavors young progenitor models, instead pointing to scenarios involving globular clusters, late-stage magnetars, or low-mass X-ray binaries (LMXBs). Though deep X-ray or polarimetric observations are not available, the localization of the FRB and a possible periodicity points to progenitors likely to be a binary involving a compact object and a stellar companion or a precessing/rotating old neutron star.
\end{abstract}


\section{Introduction} \label{sec:intro}
Fast Radio Bursts (FRBs; \citealp{2007Sci...318..777L}) are mysterious astronomical events that are defined by radio flashes that last milliseconds and come from extragalactic origins \citep{https://doi.org/10.3847/2041-8213/abdb38, arXiv:2106.09710}. Repeating FRBs \citep{2016Natur.531..202S}, which release several bursts over time, and non-repeating FRBs, which have been seen as solitary, single events, are the two different classifications into which these bursts come. The physical processes that underlie FRB emission are still unknown. There are currently $\sim 60$ repeating and $\sim 800$ \footnote{\url{https://www.wis-tns.org/}} non-repeating FRBs known to exist. FRB 121102 \citep{2014ApJ...790..101S} and FRB 180916B \citep{https://arxiv.org/pdf/1908.03507} are the only two sources in the repeating population that show periodic behavior. While FRB 180916B exhibits a shorter periodic pattern with bursts occurring every 16 days \citep{2020Natur.582..351C}, FRB 121102 exhibits an activity cycle of 160 days \citep{https://arxiv.org/pdf/2003.03596}. Although the majority of repeating FRBs do not display clear periodicity, the existence of these two periodic sources challenges theoretical models that propose completely stochastic burst generation mechanisms. Using isolated compact object mechanisms to generate periodicity on timescales of weeks to months is very difficult \citep{2020Natur.582..351C}. Two primary alternative theories have surfaced to satisfy this theoretical constraint: precession of highly magnetic neutron stars undergoing periodic flaring episodes \citep{https://ui.adsabs.harvard.edu/abs/2020arXiv200204595L,https://ui.adsabs.harvard.edu/abs/2020arXiv200205752Z} and binary orbital motion \citep{2020ApJ...893L..39L,2020ApJ...893L..26I}. Month-scale periodicity detection and constraint necessitate long-term, intensive observational follow-up.

With its four cylindrical reflector arrays offering a wide field of view of 200 square degrees \citep{2018ApJ...863...48C}, the Canadian Hydrogen Intensity Mapping Experiment (CHIME; \citealp{2018ApJ...863...48C}) is in a unique position to accomplish this objective. Its ability to continuously watch the sky guarantees thorough detection of burst activity beyond a fluence threshold of 1 Jyms \citep{2018ApJ...863...48C}, which makes it the perfect tool for both finding new FRBs and carrying out objective, long-term monitoring of established sources.

FRB 20240209A was initially detected by the CHIME-FRB collaboration \citep{2024ATel16670....1S}. Subsequent baseband observations with CHIME and the outrigger KKO station \citep{2024AJ....168...87L} revealed that FRB 20240209A is localized to the peripheral regions of a quiescent galaxy at redshift z = 0.1384 \citep{https://arxiv.org/pdf/2410.23374}. The offset of the burst from the host galactic center is significant, measuring $40 \pm 5$ kpc (68\% confidence), and the host-normalized offset is $5.13 \pm 0.6 \text{R}_{\text{eff}}$ \citep{https://arxiv.org/pdf/2410.23374}. FRB 20240209A differs from the typical FRB population, which originates primarily from star-forming galaxies with offsets of $\leq$ 10 kpc \citep{https://arxiv.org/abs/2009.10747,https://arxiv.org/abs/2012.11617,2022AJ....163...69B,https://arxiv.org/abs/2302.05465,https://arxiv.org/abs/2409.16964}, due to its remarkable galactic offset and its association with a luminous, quiescent host galaxy. Notably, this FRB has the second-highest host-normalized offset and the largest absolute host-galaxy offset, only surpassed by FRB 20200120E, which was located to a globular cluster in M81 \citep{https://arxiv.org/abs/2103.01295,https://arxiv.org/abs/2105.11445}. Furthermore, it represents the first clear association of an FRB with an elliptical galaxy \citep{https://arxiv.org/pdf/2410.23374}. The burst characteristics include a dispersion measure of 176.57(3) pc cm$^{-3}$ and a rotation measure (RM) of $<10  rad m^{-2}$ for the two reported CHIME bursts \citep{2024ATel16670....1S}.

In this paper, we make use of the publicly available CHIME-FRB repeater database, and report the discovery of a possible periodicity in the activity of the FRB 20240209A. We organize the paper as follows, we report the burst database and periodicity search in section \ref{BD} describing the core results and methods. We conclude discussing the implications over the possible progenitor models in section \ref{discussion}.

\begin{figure*}[ht!]
    \plotone{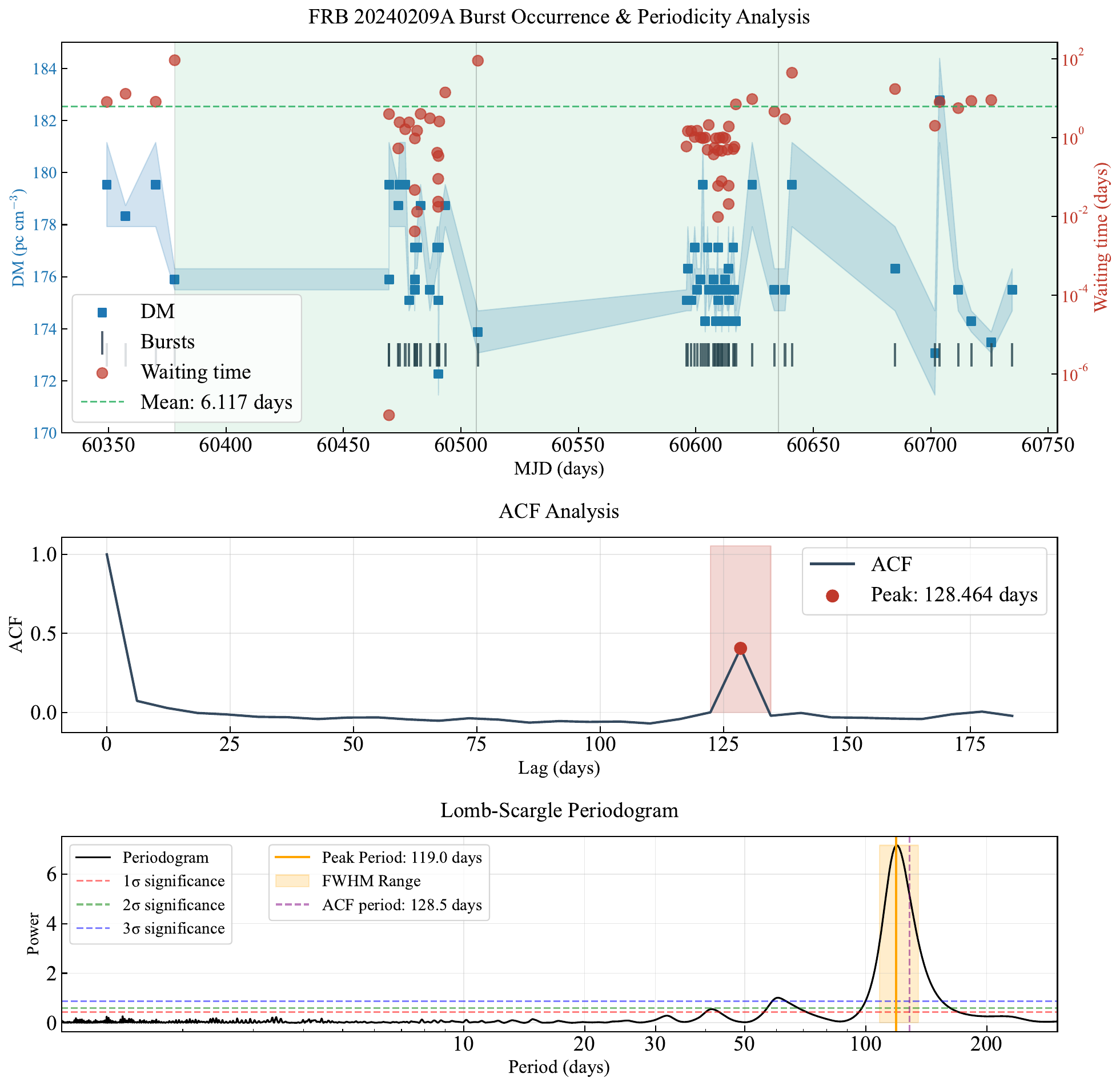}
    \caption{Temporal and periodicity analysis of FRB 20240209A. Top panel: Burst occurrence timeline showing the DM variations (blue squares), individual bursts (black ticks), and waiting times between consecutive bursts (red circles) as a function of Modified Julian Date (MJD). The mean waiting time of 5.886 days is indicated by the green dashed line. The filled blue regions show the error bars on the detected DM. The shaded green region shows the measured period visually and each box has a width of the measured ACF periodicity. Middle panel: Auto-correlation function (ACF) analysis revealing a significant peak at 123.615 days (red point with pink shaded uncertainty region). Bottom panel: Lomb-Scargle periodogram showing a strong peak at 126.6 days, with 1$\sigma$, 2$\sigma$, and 3$\sigma$ significance levels indicated by horizontal dashed lines. The FWHM range of the peak (yellow shaded region) and the ACF-derived period (purple dashed line) demonstrate the consistency between different periodicity analysis methods.}
    \label{fig:info}
\end{figure*}
\section{Burst Dataset and Periodicity Search}\label{BD}
The repeating FRB is continuously monitored by CHIME, with burst detections, arrival times, and DM measurements regularly updated in the CHIME-FRB repeaters' public database (\url{https://www.chime-frb.ca/repeaters}). Using the database, we conducted a systematic search for periodicity in the burst activity of FRB 20240209A.

On March 14, 2025, we downloaded the CHIME-FRB VOE services' repeaters' data, this serves as the foundation of our analysis. To look for periodicity, we implemented the autocorrelation function (ACF) analysis in the time series. The normalized ACF was computed by correlating the signal with itself at increasing time lags, with peaks in the resulting function corresponding to potential periodic components. The burst's ACF analysis showed a notable peak at $128.5 \pm 6$ days, suggesting that the burst activity may be recurrent. The normalized ACF indicated a relatively strong periodic signal, with this peak showing a prominence of 0.4 (Fig. \ref{fig:info}).

To confirm, we also performed a Lomb-Scargle \citep{1982ApJ...263..835S} periodogram analysis for the available bursts of the FRB 20240209A. The Lomb-Scargle periodogram performs a least-squares fit of sinusoids to the data, evaluating the power at each test frequency. Unlike simpler methods, it's specially designed to handle gaps or irregular timing in the measurements \citep{1982ApJ...263..835S}. The periodogram was computed over a frequency range corresponding to periods between 1 and 300 days. We produced random time-stamps between the days of the first and last detection of the FRB and from the times when the FRB was situated inside the CHIME field of view. We first selected random days inside the first and last detection of the FRB. We then calculated the transit time and periods on those specific dates using the CHIME-FRB transit calculator \footnote{\url https://www.chime-frb.ca/astronomytools}. The list was supplied to the bootstrapping script which does random pickups from the times when the FRB was up in the CHIME sky. Also, when the script picks dates which are not exactly in the dictionary, the transit time was corrected approximately using the sideral correction factor of 3 minutes and 56 seconds. Now with random 10,000 trials of the time stamps, bootstrap analysis was used to determine statistical significance, with confidence levels set at 1$\sigma$ (68.27\%), 2$\sigma$ (95.45\%), and 3$\sigma$ (99.73\%). At $119 \pm 13$ days, the analysis showed a dominant peak that was higher than the 3$\sigma$ confidence level (power $\geq$ 0.79).

With the ACF and Lomb-Scargle values agreeing within their uncertainties, the period obtained using both approaches demonstrates excellent consistency. Clusters of strong activity interspersed with comparatively quiet periods are how this period shows up in the burst arrival times. Our dataset shows three primary active phases, with MJD 60360, 60478, and 60630 in their centers.

Further validation of the periodic behaviour was performed through epoch folding analysis (Similar method introduced in \citealp{1978ApJ...224..953S} and following the methods described in \citealp{1983ApJ...266..160L}), which yielded a peak at P = $128.8 \pm 14$ days. Epoch folding in FRB periodicity analysis involves dividing potential periods into phase bins and
measuring how bursts cluster when ``folded" at different trial periods. When the correct period is
found, bursts cluster in specific phase bins rather than being uniformly distributed. The Pearson
chi-square test quantifies this non-uniformity by comparing observed burst counts ($N_i$) in
each phase bin against expected counts ($E_i$) under uniform distribution (Eq. \ref{pt})
A higher chi-square value indicates a periodic signal, with the highest value typically corresponding to the true period.
\begin{align}
\chi^2 = \sum_{i=1}^{n} \frac{(N_i - E_i)^2}{E_i}
\label{pt}
\end{align}

\begin{figure*}[ht!]
    \epsscale{0.8}
    \plotone{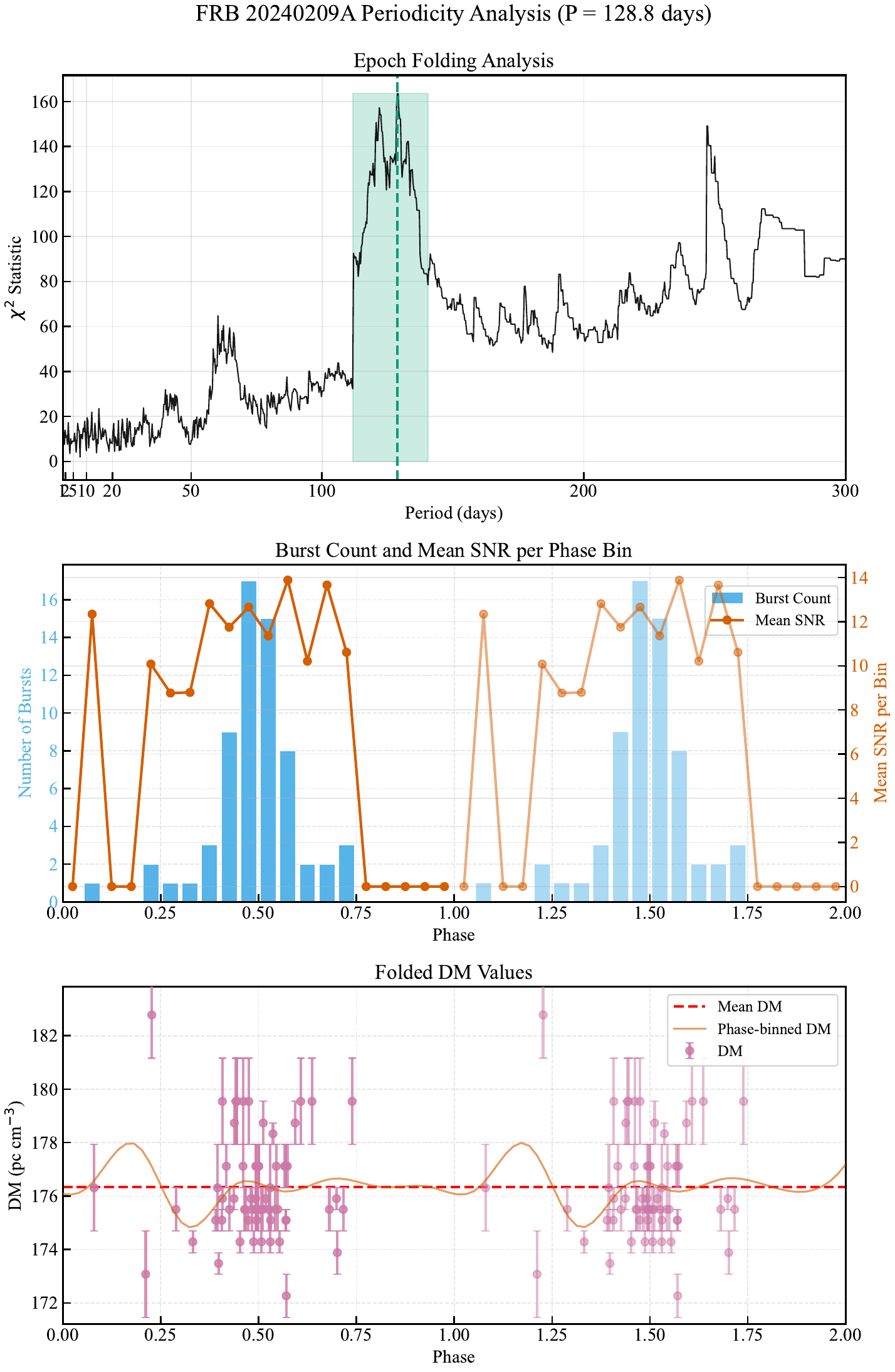}
    \caption{Periodicity analysis of FRB 20240209A using the epoch folding method. Top panel: The $\chi^2$ statistic as a function of trial periods, showing a significant peak at $128.8 \pm 15.4$ days (marked by the green dashed line with green shaded region indicating the uncertainty). Middle panel: The total burst count distribution folded at the best-fit period of 128.8 days is shown in blue. The mean SNR across every phase interval is shown in a red line connecting the red dots. Bottom panel: Evolution of the DM after folding with the period of 128.8 days is shown in violet circles, with the phase averaged DM is shown as orange line. The average DM is shown in dashed red line.}
    \label{fig:epoch-folding}
\end{figure*}

The uncertainty is derived from the full width at half maximum (FWHM) of the $\chi^{2}$ peak (shaded region in Fig. \ref{fig:epoch-folding}). The epoch folding method splits the time series into phase bins using trial periods, computing a $\chi^{2}$ statistic to measure deviations from uniform distribution. The peak period agrees with both Lomb-Scargle ($119 \pm 13$ days) and ACF ($128.5 \pm 6$ days) results within the uncertainties. We also have folded the profile using the periodicity of 128.8 days and the folded profiles shows primary concentrations starting at phases $\phi \sim 0.0 $ and $\phi \sim 0.9$ and a noticeable activity gap between $\phi \sim 0.3-0.6$ and from $\sim 0.65-0.9$. The folded profile shows the highest number of bursts in the beginning and the end of the activity phase. 

From the Tab. \ref{tab:frb_bursts}, the bursts have different detection significances which is unlikely to significantly impact the periodicity measurements. We do not have any estimates on the burst fluence from the CHIME VOE database but most of the bursts were reported on \cite{https://arxiv.org/pdf/2410.23374} where the bursts fluence ranges from 7.4 - 324.9 Jy.ms. The CHIME FRB has a fluence complete limit of 1 Jy.ms \cite{2018ApJ...863...48C}. Though, the possibility of missing a broad fainter population can not be ruled out, the burst activity would still be periodic in terms of energy scales.
 
\section{Implications on the Progenitor Models}\label{discussion}
FRB 20240209A has been proposed to be possibly associated with a globular cluster in its host galaxy \citep{https://arxiv.org/pdf/2410.23374}. The observed host dispersion measure and comparison of offsets with soft-gamma-ray repeaters make alternative scenarios, such as an undetected dwarf galaxy or a kicked-off progenitor, less plausible \citep{https://arxiv.org/pdf/2410.23374}. Progenitor models incorporating elderly star populations, such as magnetars generated by accretion-induced collapse of white dwarfs (WDs), merger-induced collapse of WD-WD binaries, NS-WD systems, or NS-NS binaries, are especially favored in the globular cluster environment \citep{https://arxiv.org/pdf/2410.23374}.

Given their demonstrated prevalence in globular clusters, low-mass X-ray binaries (LMXBs) and ultraluminous X-ray sources (ULXs) \citep{1975ApJ...199L.143C,https://arxiv.org/pdf/astro-ph/0701310, https://arxiv.org/abs/2103.16576,https://arxiv.org/abs/2102.06138} are other promising candidates. The periodic character of the burst activity can suggest of a binary system. The 126 days periodicity disfavours binaries with both compact objects due to expected shorter time scales \citep{2014LRR....17....3P}. It may hint towards a binary system involving a compact object and a star partner. Either the compact object itself or the interaction of magnetic fields with material stripped from the companion star could be the source of FRB emission in such situations.\citep{https://arxiv.org/abs/2102.06138,2020ApJ...893L..39L,2020ApJ...893L..26I}

Other possibilities also remain open as the periodicity can be attributed to the precession \citep{https://ui.adsabs.harvard.edu/abs/2020arXiv200204595L,https://ui.adsabs.harvard.edu/abs/2020arXiv200205752Z} or rotational motion of old stage magnetars \citep{https://arxiv.org/pdf/2003.12509,https://ui.adsabs.harvard.edu/abs/2021ApJ...917....2X/abstract,https://arxiv.org/pdf/2210.09323}. The nature of this source will need to be further constrained by future in-depth polarimetric investigations, particularly including measurements of position angle variations. For both the binary and precessing models, a shorter time-scale periodicity is also expected due to the spin of the magnetar itself \citep{https://arxiv.org/pdf/2210.09323} but currently the limited daily exposure with CHIME can not constrain a periodicity of more than 15-25 minutes (assuming daily exposure of 74 minutes). Facilities with no bound on the exposure time can be well suited to look for these short scale periods in the next expected burst activity window. If the FRBs are getting generated through interactions of the ablated material and the motion of the magnetar, a flat or slowly varying polarization angle (PA) swing can be expected if the magnetic field configuration does not change with time. However, in the case of precession and rotation, the PA will have an imprint on them. For the precession, if the smaller scale periodicity is found, the PA is expected to exhibit a swing in the magnetar's repeating timescale. Same holds for the rotation but in this we can expect a much slower and gradual swing, 128 days being attributed to the periods of the magnetar.

Later epochs (Fig. \ref{fig:info}) reveal increased burst activity for FRB 20240209A, with the most recent events in January and February 2025 showing a decline in activity. This pattern may indicate that FRB formation is driven by interactions between ablated material and a compact object in a binary system. This variation in burst rate could be explained by changes in the companion wind density or velocity, which affects the number of particles entering the magnetosphere of the neutron star through the funnel interface \citep{https://arxiv.org/pdf/2105.14480}. According to the binary comb model \citep{2020ApJ...893L..26I,https://arxiv.org/pdf/2105.14480}, these variations could create aurora-like particle flows that alter emission characteristics without requiring rapid orbital evolution. By continuing to observe the source's periodic behavior over an extended period of time, we may detect changes in dispersion measure that would provide a stronger signature of the binary interaction.

The FRB's low reported rotation measure (RM) indicates that it originated in or propagated through a turbulent medium. Comprehensive RM information about this source is currently missing from the CHIME-repeaters database. Important information about the local environment and the underlying progenitor mechanism may be revealed by a thorough analysis of RM variations across detected bursts \citep{https://arxiv.org/abs/2204.08124}.

The dispersion measure (DM), available through the database, exhibits periodic modulation within the activity windows. While the magnetar-Be star binary model \citep{https://arxiv.org/abs/2204.08124}, which successfully explains the RM variations in FRB 20201124A with its characteristic zig-zag pattern, predicts a monotonic increase followed by a decrease in DM (Fig. 3 in \citealp{https://arxiv.org/abs/2204.08124}). The FRB 20240209A displays distinct oscillatory behavior.
Notably, as seen in Fig \ref{fig:info} (top panel), the DM exhibits a similar modulation pattern throughout all activity phases. A decreased DM value starts the pattern, which then follows an increase, then a reduction, and then several periodic modulations (especially in the second and third activity phases), and eventually a decline. The electron density structure of the ablated material in the disk of the binary system may be traced by this behavior. The period folded DM also shows distinct oscillatory behaviour (Fig. \ref{fig:epoch-folding} Bottom Panel).
Since interactions between the compact object and its partner are likely to cause persistent X-ray emission in such systems, deep X-ray follow-up observations will be essential for constraining the binary progenitor concept \citep{https://arxiv.org/abs/2105.11445}. But the even for the current state of the art telescopes like Chandra, 1 Ms is needed to reach $10^{39}$ ergs s$^{-1}$, while that can constrain the progenitor if it is a Ultraluminous Xray Binary \citep{2005ApJ...624L..17S,2025NatAs...9..111P} but way above if the progenitor is an late-stage magnetars ($10^{29} - 10^{36}$ ergs s$^{-1}$; \citealp{2018MNRAS.474..961C}), extragalactic millisecond pulsars ($10^{34} - 10^{37}$ ergs s$^{-1}$; \citealp{2018ApJ...861....5K,2021ApJ...908..212G}), and LMXBs ($10^{30} - 10^{39}$ ergs s$^{-1}$;  \citealp{2025NatAs...9..111P} ).

The FRB position was monitored by the CHIME/FRB for quite a long time before the discovery of the FRB in 2024. The sudden activation can be possibly attributed to magnetic field reconfiguration in an old magnetar or the onset of interactions in a binary system that reached a critical threshold. In the binary scenario, material ablation from the companion star may have only recently begun feeding the magnetosphere of the compact object, triggering the FRB mechanism. Alternatively, in a precessing magnetar model, the emission beam may have only recently aligned with our line of sight after years of pointing elsewhere.

The localized position of FRB 20240209A was continuously monitored by CHIME/FRB for several years before its sudden discovery in 2024. This abrupt activation can be potentially explained through multiple progenitor models. In the binary scenario, the system may have reached a critical orbital configuration where material ablation from the companion star began feeding the magnetosphere of the compact object, triggering the FRB mechanism \citep{2020ApJ...893L..26I,2020Natur.587...45Z}. If the source is a rotating or precessing old-stage magnetar, the emission beam may have only recently aligned with our line of sight. Or, the sudden "switch-on" could represent a magnetic field reconfiguration event in the magnetar\citep{2015ApJ...800...76S,2020MNRAS.498..651B}, where rapid crustal failures or magnetospheric instabilities trigger coherent radio emission that was previously absent. These magnetic reconfigurations can occur stochastically or be triggered by external perturbations. A single large-scale rearrangement of the internal magnetic field can initiate a large-angle free precession \citep{2020ApJ...895L..30L}.

The discovered possible $\sim$ 4-month periodicity and the FRB's substantial offset from its host galaxy provide compelling evidence for a binary origin of the progenitor, most likely comprising a compact object-stellar companion system. Though the currently available data can not differentiate surely between the other possibilities being a old stage rotating/precessing magnetar. The data only comprises of 2 full periods, which is just sufficient to probe any periodicity. Hence, comprehensive radio monitoring campaigns remain essential to definitively confirm the periodicity, especially with polarization to weigh all the progenitor models and constrain.

\section{Acknowledgements}
We acknowledge use of the CHIME/FRB Public Database, provided at https://www.chime-frb.ca/ by the CHIME/FRB Collaboration. This research has made use of NASA's Astrophysics Data System Bibliographic Services. This research has made use of the NASA/IPAC Extragalactic Database (NED),
which is operated by the Jet Propulsion Laboratory, California Institute of Technology,
under contract with the National Aeronautics and Space Administration.
\section{Data availability}
The data is publicly available on \href{http://www.chime-frb.ca/repeaters}{CHIME repeaters' database}. 

% \section{Code availability}
% The code can be found in \url{https://github.com/arpan-52/frb-2024}
\bibliographystyle{aasjournal}
\bibliography{frb}

\begin{thebibliography}{}
\expandafter\ifx\csname natexlab\endcsname\relax\def\natexlab#1{#1}\fi
\providecommand{\url}[1]{\href{#1}{#1}}
\providecommand{\dodoi}[1]{doi:~\href{http://doi.org/#1}{\nolinkurl{#1}}}
\providecommand{\doeprint}[1]{\href{http://ascl.net/#1}{\nolinkurl{http://ascl.net/#1}}}
\providecommand{\doarXiv}[1]{\href{https://arxiv.org/abs/#1}{\nolinkurl{https://arxiv.org/abs/#1}}}

\bibitem[{{Beniamini} \& {Kumar}(2020)}]{2020MNRAS.498..651B}
{Beniamini}, P., \& {Kumar}, P. 2020, \mnras, 498, 651, \dodoi{10.1093/mnras/staa2489}

\bibitem[{{Beniamini} {et~al.}(2023){Beniamini}, {Wadiasingh}, {Hare}, {Rajwade}, {Younes}, \& {van der Horst}}]{https://arxiv.org/pdf/2210.09323}
{Beniamini}, P., {Wadiasingh}, Z., {Hare}, J., {et~al.} 2023, \mnras, 520, 1872, \dodoi{10.1093/mnras/stad208}

\bibitem[{{Beniamini} {et~al.}(2020){Beniamini}, {Wadiasingh}, \& {Metzger}}]{https://arxiv.org/pdf/2003.12509}
{Beniamini}, P., {Wadiasingh}, Z., \& {Metzger}, B.~D. 2020, \mnras, 496, 3390, \dodoi{10.1093/mnras/staa1783}

\bibitem[{{Bhandari} {et~al.}(2022){Bhandari}, {Heintz}, {Aggarwal}, {Marnoch}, {Day}, {Sydnor}, {Burke-Spolaor}, {Law}, {Xavier Prochaska}, {Tejos}, {Bannister}, {Butler}, {Deller}, {Ekers}, {Flynn}, {Fong}, {James}, {Lazio}, {Luo}, {Mahony}, {Ryder}, {Sadler}, {Shannon}, {Han}, {Lee}, \& {Zhang}}]{2022AJ....163...69B}
{Bhandari}, S., {Heintz}, K.~E., {Aggarwal}, K., {et~al.} 2022, \aj, 163, 69, \dodoi{10.3847/1538-3881/ac3aec}

\bibitem[{{Bhardwaj} {et~al.}(2021){Bhardwaj}, {Gaensler}, {Kaspi}, {Landecker}, {Mckinven}, {Michilli}, {Pleunis}, {Tendulkar}, {Andersen}, {Boyle}, {Cassanelli}, {Chawla}, {Cook}, {Dobbs}, {Fonseca}, {Kaczmarek}, {Leung}, {Masui}, {Mnchmeyer}, {Ng}, {Rafiei-Ravandi}, {Scholz}, {Shin}, {Smith}, {Stairs}, \& {Zwaniga}}]{https://arxiv.org/abs/2103.01295}
{Bhardwaj}, M., {Gaensler}, B.~M., {Kaspi}, V.~M., {et~al.} 2021, \apjl, 910, L18, \dodoi{10.3847/2041-8213/abeaa6}

\bibitem[{{CHIME/FRB Collaboration} {et~al.}(2018){CHIME/FRB Collaboration}, {Amiri}, {Bandura}, {Berger}, {Bhardwaj}, {Boyce}, {Boyle}, {Brar}, {Burhanpurkar}, {Chawla}, {Chowdhury}, {Cliche}, {Cranmer}, {Cubranic}, {Deng}, {Denman}, {Dobbs}, {Fandino}, {Fonseca}, {Gaensler}, {Giri}, {Gilbert}, {Good}, {Guliani}, {Halpern}, {Hinshaw}, {H{\"o}fer}, {Josephy}, {Kaspi}, {Landecker}, {Lang}, {Liao}, {Masui}, {Mena-Parra}, {Naidu}, {Newburgh}, {Ng}, {Patel}, {Pen}, {Pinsonneault-Marotte}, {Pleunis}, {Rafiei Ravandi}, {Ransom}, {Renard}, {Scholz}, {Sigurdson}, {Siegel}, {Smith}, {Stairs}, {Tendulkar}, {Vanderlinde}, \& {Wiebe}}]{2018ApJ...863...48C}
{CHIME/FRB Collaboration}, {Amiri}, M., {Bandura}, K., {et~al.} 2018, \apj, 863, 48, \dodoi{10.3847/1538-4357/aad188}

\bibitem[{{CHIME/FRB Collaboration} {et~al.}(2019){CHIME/FRB Collaboration}, {Andersen}, {Bandura}, {Bhardwaj}, {Boubel}, {Boyce}, {Boyle}, {Brar}, {Cassanelli}, {Chawla}, {Cubranic}, {Deng}, {Dobbs}, {Fandino}, {Fonseca}, {Gaensler}, {Gilbert}, {Giri}, {Good}, {Halpern}, {Hill}, {Hinshaw}, {H{\"o}fer}, {Josephy}, {Kaspi}, {Kothes}, {Landecker}, {Lang}, {Li}, {Lin}, {Masui}, {Mena-Parra}, {Merryfield}, {Mckinven}, {Michilli}, {Milutinovic}, {Naidu}, {Newburgh}, {Ng}, {Patel}, {Pen}, {Pinsonneault-Marotte}, {Pleunis}, {Rafiei-Ravandi}, {Rahman}, {Ransom}, {Renard}, {Scholz}, {Siegel}, {Singh}, {Smith}, {Stairs}, {Tendulkar}, {Tretyakov}, {Vanderlinde}, {Yadav}, \& {Zwaniga}}]{https://arxiv.org/pdf/1908.03507}
{CHIME/FRB Collaboration}, {Andersen}, B.~C., {Bandura}, K., {et~al.} 2019, \apjl, 885, L24, \dodoi{10.3847/2041-8213/ab4a80}

\bibitem[{{Chime/Frb Collaboration} {et~al.}(2020){Chime/Frb Collaboration}, {Amiri}, {Andersen}, {Bandura}, {Bhardwaj}, {Boyle}, {Brar}, {Chawla}, {Chen}, {Cliche}, {Cubranic}, {Deng}, {Denman}, {Dobbs}, {Dong}, {Fandino}, {Fonseca}, {Gaensler}, {Giri}, {Good}, {Halpern}, {Hessels}, {Hill}, {H{\"o}fer}, {Josephy}, {Kania}, {Karuppusamy}, {Kaspi}, {Keimpema}, {Kirsten}, {Landecker}, {Lang}, {Leung}, {Li}, {Lin}, {Marcote}, {Masui}, {McKinven}, {Mena-Parra}, {Merryfield}, {Michilli}, {Milutinovic}, {Mirhosseini}, {Naidu}, {Newburgh}, {Ng}, {Nimmo}, {Paragi}, {Patel}, {Pen}, {Pinsonneault-Marotte}, {Pleunis}, {Rafiei-Ravandi}, {Rahman}, {Ransom}, {Renard}, {Sanghavi}, {Scholz}, {Shaw}, {Shin}, {Siegel}, {Singh}, {Smegal}, {Smith}, {Stairs}, {Tendulkar}, {Tretyakov}, {Vanderlinde}, {Wang}, {Wang}, {Wulf}, {Yadav}, \& {Zwaniga}}]{2020Natur.582..351C}
{Chime/Frb Collaboration}, {Amiri}, M., {Andersen}, B.~C., {et~al.} 2020, \nat, 582, 351, \dodoi{10.1038/s41586-020-2398-2}

\bibitem[{{Clark}(1975)}]{1975ApJ...199L.143C}
{Clark}, G.~W. 1975, \apjl, 199, L143, \dodoi{10.1086/181869}

\bibitem[{{Coti Zelati} {et~al.}(2018){Coti Zelati}, {Rea}, {Pons}, {Campana}, \& {Esposito}}]{2018MNRAS.474..961C}
{Coti Zelati}, F., {Rea}, N., {Pons}, J.~A., {Campana}, S., \& {Esposito}, P. 2018, \mnras, 474, 961, \dodoi{10.1093/mnras/stx2679}

\bibitem[{{Dage} {et~al.}(2021){Dage}, {Kundu}, {Thygesen}, {Bahramian}, {Haggard}, {Irwin}, {Maccarone}, {Nair}, {Peacock}, {Strader}, \& {Zepf}}]{https://arxiv.org/abs/2103.16576}
{Dage}, K.~C., {Kundu}, A., {Thygesen}, E., {et~al.} 2021, \mnras, 504, 1545, \dodoi{10.1093/mnras/stab943}

\bibitem[{{Gordon} {et~al.}(2023){Gordon}, {Fong}, {Kilpatrick}, {Eftekhari}, {Leja}, {Prochaska}, {Nugent}, {Bhandari}, {Blanchard}, {Caleb}, {Day}, {Deller}, {Dong}, {Glowacki}, {Gourdji}, {Mannings}, {Mahoney}, {Marnoch}, {Miller}, {Paterson}, {Rastinejad}, {Ryder}, {Sadler}, {Scott}, {Sears}, {Shannon}, {Simha}, {Stappers}, \& {Tejos}}]{https://arxiv.org/abs/2302.05465}
{Gordon}, A.~C., {Fong}, W.-f., {Kilpatrick}, C.~D., {et~al.} 2023, \apj, 954, 80, \dodoi{10.3847/1538-4357/ace5aa}

\bibitem[{{Gotthelf} {et~al.}(2021){Gotthelf}, {Safi-Harb}, {Straal}, \& {Gelfand}}]{2021ApJ...908..212G}
{Gotthelf}, E.~V., {Safi-Harb}, S., {Straal}, S.~M., \& {Gelfand}, J.~D. 2021, \apj, 908, 212, \dodoi{10.3847/1538-4357/abd32b}

\bibitem[{{Heintz} {et~al.}(2020){Heintz}, {Prochaska}, {Simha}, {Platts}, {Fong}, {Tejos}, {Ryder}, {Aggerwal}, {Bhandari}, {Day}, {Deller}, {Kilpatrick}, {Law}, {Macquart}, {Mannings}, {Marnoch}, {Sadler}, \& {Shannon}}]{https://arxiv.org/abs/2009.10747}
{Heintz}, K.~E., {Prochaska}, J.~X., {Simha}, S., {et~al.} 2020, \apj, 903, 152, \dodoi{10.3847/1538-4357/abb6fb}

\bibitem[{{Ioka} \& {Zhang}(2020)}]{2020ApJ...893L..26I}
{Ioka}, K., \& {Zhang}, B. 2020, \apjl, 893, L26, \dodoi{10.3847/2041-8213/ab83fb}

\bibitem[{{Kirsten} {et~al.}(2022){Kirsten}, {Marcote}, {Nimmo}, {Hessels}, {Bhardwaj}, {Tendulkar}, {Keimpema}, {Yang}, {Snelders}, {Scholz}, {Pearlman}, {Law}, {Peters}, {Giroletti}, {Paragi}, {Bassa}, {Hewitt}, {Bach}, {Bezrukovs}, {Burgay}, {Buttaccio}, {Conway}, {Corongiu}, {Feiler}, {Forss{\'e}n}, {Gawro{\'n}ski}, {Karuppusamy}, {Kharinov}, {Lindqvist}, {Maccaferri}, {Melnikov}, {Ould-Boukattine}, {Possenti}, {Surcis}, {Wang}, {Yuan}, {Aggarwal}, {Anna-Thomas}, {Bower}, {Blaauw}, {Burke-Spolaor}, {Cassanelli}, {Clarke}, {Fonseca}, {Gaensler}, {Gopinath}, {Kaspi}, {Kassim}, {Lazio}, {Leung}, {Li}, {Lin}, {Masui}, {Mckinven}, {Michilli}, {Mikhailov}, {Ng}, {Orbidans}, {Pen}, {Petroff}, {Rahman}, {Ransom}, {Shin}, {Smith}, {Stairs}, \& {Vlemmings}}]{https://arxiv.org/abs/2105.11445}
{Kirsten}, F., {Marcote}, B., {Nimmo}, K., {et~al.} 2022, \nat, 602, 585, \dodoi{10.1038/s41586-021-04354-w}

\bibitem[{{Klingler} {et~al.}(2018){Klingler}, {Kargaltsev}, {Pavlov}, {Ng}, {Beniamini}, \& {Volkov}}]{2018ApJ...861....5K}
{Klingler}, N., {Kargaltsev}, O., {Pavlov}, G.~G., {et~al.} 2018, \apj, 861, 5, \dodoi{10.3847/1538-4357/aac6e0}

\bibitem[{{Lanman} {et~al.}(2024){Lanman}, {Andrew}, {Lazda}, {Shah}, {Amiri}, {Balasubramanian}, {Bandura}, {Boyle}, {Brar}, {Carlson}, {Cliche}, {Gusinskaia}, {Hendricksen}, {Kaczmarek}, {Landecker}, {Leung}, {Mckinven}, {Mena-Parra}, {Milutinovic}, {Nimmo}, {Pearlman}, {Renard}, {Rahman}, {Shaw}, {Siegel}, {Smegal}, {Cassanelli}, {Chatterjee}, {Curtin}, {Dobbs}, {Dong}, {Halpern}, {Hopkins}, {Kaspi}, {Khairy}, {Masui}, {Meyers}, {Michilli}, {Petroff}, {Pinsonneault-Marotte}, {Pleunis}, {Rafiei-Ravandi}, {Shin}, {Smith}, {Vanderlinde}, \& {Zegmott}}]{2024AJ....168...87L}
{Lanman}, A.~E., {Andrew}, S., {Lazda}, M., {et~al.} 2024, \aj, 168, 87, \dodoi{10.3847/1538-3881/ad5838}

\bibitem[{{Law} {et~al.}(2024){Law}, {Bhardwaj}, {Burke-Spolaor}, {Thomas}, {Demorest}, \& {Bower}}]{2024ATel16701....1L}
{Law}, C.~J., {Bhardwaj}, M., {Burke-Spolaor}, S., {et~al.} 2024, The Astronomer's Telegram, 16701, 1

\bibitem[{{Leahy} {et~al.}(1983){Leahy}, {Darbro}, {Elsner}, {Weisskopf}, {Sutherland}, {Kahn}, \& {Grindlay}}]{1983ApJ...266..160L}
{Leahy}, D.~A., {Darbro}, W., {Elsner}, R.~F., {et~al.} 1983, \apj, 266, 160, \dodoi{10.1086/160766}

\bibitem[{{Levin} {et~al.}(2020{\natexlab{a}}){Levin}, {Beloborodov}, \& {Bransgrove}}]{https://ui.adsabs.harvard.edu/abs/2020arXiv200204595L}
{Levin}, Y., {Beloborodov}, A.~M., \& {Bransgrove}, A. 2020{\natexlab{a}}, \apjl, 895, L30, \dodoi{10.3847/2041-8213/ab8c4c}

\bibitem[{{Levin} {et~al.}(2020{\natexlab{b}}){Levin}, {Beloborodov}, \& {Bransgrove}}]{2020ApJ...895L..30L}
---. 2020{\natexlab{b}}, \apjl, 895, L30, \dodoi{10.3847/2041-8213/ab8c4c}

\bibitem[{{Lorimer} {et~al.}(2007){Lorimer}, {Bailes}, {McLaughlin}, {Narkevic}, \& {Crawford}}]{2007Sci...318..777L}
{Lorimer}, D.~R., {Bailes}, M., {McLaughlin}, M.~A., {Narkevic}, D.~J., \& {Crawford}, F. 2007, Science, 318, 777, \dodoi{10.1126/science.1147532}

\bibitem[{{Lyutikov} {et~al.}(2020){Lyutikov}, {Barkov}, \& {Giannios}}]{2020ApJ...893L..39L}
{Lyutikov}, M., {Barkov}, M.~V., \& {Giannios}, D. 2020, \apjl, 893, L39, \dodoi{10.3847/2041-8213/ab87a4}

\bibitem[{{Maccarone} {et~al.}(2007){Maccarone}, {Kundu}, {Zepf}, \& {Rhode}}]{https://arxiv.org/pdf/astro-ph/0701310}
{Maccarone}, T.~J., {Kundu}, A., {Zepf}, S.~E., \& {Rhode}, K.~L. 2007, \nat, 445, 183, \dodoi{10.1038/nature05434}

\bibitem[{{Mannings} {et~al.}(2021){Mannings}, {Fong}, {Simha}, {Prochaska}, {Rafelski}, {Kilpatrick}, {Tejos}, {Heintz}, {Bannister}, {Bhandari}, {Day}, {Deller}, {Ryder}, {Shannon}, \& {Tendulkar}}]{https://arxiv.org/abs/2012.11617}
{Mannings}, A.~G., {Fong}, W.-f., {Simha}, S., {et~al.} 2021, \apj, 917, 75, \dodoi{10.3847/1538-4357/abff56}

\bibitem[{{Ould-Boukattine} {et~al.}(2024){Ould-Boukattine}, {Hessels}, {Snelders}, {Kirsten}, {Blaauw}, {Sluman}, \& {Mulder}}]{2024ATel16732....1O}
{Ould-Boukattine}, O.~S., {Hessels}, J.~W.~T., {Snelders}, M.~P., {et~al.} 2024, The Astronomer's Telegram, 16732, 1

\bibitem[{{Pearlman} {et~al.}(2025){Pearlman}, {Scholz}, {Bethapudi}, {Hessels}, {Kaspi}, {Kirsten}, {Nimmo}, {Spitler}, {Fonseca}, {Meyers}, {Stairs}, {Tan}, {Bhardwaj}, {Chatterjee}, {Cook}, {Curtin}, {Dong}, {Eftekhari}, {Gaensler}, {G{\"u}ver}, {Kaczmarek}, {Leung}, {Masui}, {Michilli}, {Prince}, {Sand}, {Shin}, {Smith}, \& {Tendulkar}}]{2025NatAs...9..111P}
{Pearlman}, A.~B., {Scholz}, P., {Bethapudi}, S., {et~al.} 2025, Nature Astronomy, 9, 111, \dodoi{10.1038/s41550-024-02386-6}

\bibitem[{{Postnov} \& {Yungelson}(2014)}]{2014LRR....17....3P}
{Postnov}, K.~A., \& {Yungelson}, L.~R. 2014, Living Reviews in Relativity, 17, 3, \dodoi{10.12942/lrr-2014-3}

\bibitem[{{Rajwade} {et~al.}(2020){Rajwade}, {Mickaliger}, {Stappers}, {Morello}, {Agarwal}, {Bassa}, {Breton}, {Caleb}, {Karastergiou}, {Keane}, \& {Lorimer}}]{https://arxiv.org/pdf/2003.03596}
{Rajwade}, K.~M., {Mickaliger}, M.~B., {Stappers}, B.~W., {et~al.} 2020, \mnras, 495, 3551, \dodoi{10.1093/mnras/staa1237}

\bibitem[{{Ravi} {et~al.}(2022){Ravi}, {Law}, {Li}, {Aggarwal}, {Bhardwaj}, {Burke-Spolaor}, {Connor}, {Lazio}, {Simard}, {Somalwar}, \& {Tendulkar}}]{arXiv:2106.09710}
{Ravi}, V., {Law}, C.~J., {Li}, D., {et~al.} 2022, \mnras, 513, 982, \dodoi{10.1093/mnras/stac465}

\bibitem[{{Scargle}(1982)}]{1982ApJ...263..835S}
{Scargle}, J.~D. 1982, \apj, 263, 835, \dodoi{10.1086/160554}

\bibitem[{{Shah} \& {CHIME/FRB Collaboration}(2024)}]{2024ATel16670....1S}
{Shah}, V., \& {CHIME/FRB Collaboration}. 2024, The Astronomer's Telegram, 16670, 1

\bibitem[{{Shah} {et~al.}(2025){Shah}, {Shin}, {Leung}, {Fong}, {Eftekhari}, {Amiri}, {Andersen}, {Andrew}, {Bhardwaj}, {Brar}, {Cassanelli}, {Chatterjee}, {Curtin}, {Dobbs}, {Dong}, {Dong}, {Fonseca}, {Gaensler}, {Halpern}, {Hessels}, {Ibik}, {Jain}, {Joseph}, {Kaczmarek}, {Kahinga}, {Kaspi}, {Kharel}, {Landecker}, {Lanman}, {Lazda}, {Main}, {Mas-Ribas}, {Masui}, {Mckinven}, {Mena-Parra}, {Meyers}, {Michilli}, {Nimmo}, {Pandhi}, {Patil}, {Pearlman}, {Pleunis}, {Prochaska}, {Rafiei-Ravandi}, {Sammons}, {Sand}, {Scholz}, {Smith}, \& {Stairs}}]{https://arxiv.org/pdf/2410.23374}
{Shah}, V., {Shin}, K., {Leung}, C., {et~al.} 2025, \apjl, 979, L21, \dodoi{10.3847/2041-8213/ad9ddc}

\bibitem[{{Sharma} {et~al.}(2024){Sharma}, {Ravi}, {Connor}, {Law}, {Ocker}, {Sherman}, {Kosogorov}, {Faber}, {Hallinan}, {Harnach}, {Hellbourg}, {Hobbs}, {Hodge}, {Hodges}, {Lamb}, {Rasmussen}, {Somalwar}, {Weinreb}, {Woody}, {Leja}, {Anand}, {Das}, {Qin}, {Rose}, {Dong}, {Miller}, \& {Yao}}]{https://arxiv.org/abs/2409.16964}
{Sharma}, K., {Ravi}, V., {Connor}, L., {et~al.} 2024, \nat, 635, 61, \dodoi{10.1038/s41586-024-08074-9}

\bibitem[{{Sivakoff} {et~al.}(2005){Sivakoff}, {Sarazin}, \& {Jord{\'a}n}}]{2005ApJ...624L..17S}
{Sivakoff}, G.~R., {Sarazin}, C.~L., \& {Jord{\'a}n}, A. 2005, \apjl, 624, L17, \dodoi{10.1086/430374}

\bibitem[{{Spitler} {et~al.}(2014){Spitler}, {Cordes}, {Hessels}, {Lorimer}, {McLaughlin}, {Chatterjee}, {Crawford}, {Deneva}, {Kaspi}, {Wharton}, {Allen}, {Bogdanov}, {Brazier}, {Camilo}, {Freire}, {Jenet}, {Karako-Argaman}, {Knispel}, {Lazarus}, {Lee}, {van Leeuwen}, {Lynch}, {Ransom}, {Scholz}, {Siemens}, {Stairs}, {Stovall}, {Swiggum}, {Venkataraman}, {Zhu}, {Aulbert}, \& {Fehrmann}}]{2014ApJ...790..101S}
{Spitler}, L.~G., {Cordes}, J.~M., {Hessels}, J.~W.~T., {et~al.} 2014, \apj, 790, 101, \dodoi{10.1088/0004-637X/790/2/101}

\bibitem[{{Spitler} {et~al.}(2016){Spitler}, {Scholz}, {Hessels}, {Bogdanov}, {Brazier}, {Camilo}, {Chatterjee}, {Cordes}, {Crawford}, {Deneva}, {Ferdman}, {Freire}, {Kaspi}, {Lazarus}, {Lynch}, {Madsen}, {McLaughlin}, {Patel}, {Ransom}, {Seymour}, {Stairs}, {Stappers}, {van Leeuwen}, \& {Zhu}}]{2016Natur.531..202S}
{Spitler}, L.~G., {Scholz}, P., {Hessels}, J.~W.~T., {et~al.} 2016, \nat, 531, 202, \dodoi{10.1038/nature17168}

\bibitem[{{Sridhar} {et~al.}(2021){Sridhar}, {Metzger}, {Beniamini}, {Margalit}, {Renzo}, {Sironi}, \& {Kovlakas}}]{https://arxiv.org/abs/2102.06138}
{Sridhar}, N., {Metzger}, B.~D., {Beniamini}, P., {et~al.} 2021, \apj, 917, 13, \dodoi{10.3847/1538-4357/ac0140}

\bibitem[{{Stellingwerf}(1978)}]{1978ApJ...224..953S}
{Stellingwerf}, R.~F. 1978, \apj, 224, 953, \dodoi{10.1086/156444}

\bibitem[{{Szary} {et~al.}(2015){Szary}, {Melikidze}, \& {Gil}}]{2015ApJ...800...76S}
{Szary}, A., {Melikidze}, G.~I., \& {Gil}, J. 2015, \apj, 800, 76, \dodoi{10.1088/0004-637X/800/1/76}

\bibitem[{{Tendulkar} {et~al.}(2021){Tendulkar}, {Gil de Paz}, {Kirichenko}, {Hessels}, {Bhardwaj}, {{\'A}vila}, {Bassa}, {Chawla}, {Fonseca}, {Kaspi}, {Keimpema}, {Kirsten}, {Lazio}, {Marcote}, {Masui}, {Nimmo}, {Paragi}, {Rahman}, {Pay{\'a}}, {Scholz}, \& {Stairs}}]{https://doi.org/10.3847/2041-8213/abdb38}
{Tendulkar}, S.~P., {Gil de Paz}, A., {Kirichenko}, A.~Y., {et~al.} 2021, \apjl, 908, L12, \dodoi{10.3847/2041-8213/abdb38}

\bibitem[{{Wada} {et~al.}(2021){Wada}, {Ioka}, \& {Zhang}}]{https://arxiv.org/pdf/2105.14480}
{Wada}, T., {Ioka}, K., \& {Zhang}, B. 2021, \apj, 920, 54, \dodoi{10.3847/1538-4357/ac127a}

\bibitem[{{Wang} {et~al.}(2022){Wang}, {Zhang}, {Dai}, \& {Cheng}}]{https://arxiv.org/abs/2204.08124}
{Wang}, F.~Y., {Zhang}, G.~Q., {Dai}, Z.~G., \& {Cheng}, K.~S. 2022, Nature Communications, 13, 4382, \dodoi{10.1038/s41467-022-31923-y}

\bibitem[{{Xu} {et~al.}(2021){Xu}, {Li}, {Yang}, {Li}, {Dai}, \& {Liu}}]{https://ui.adsabs.harvard.edu/abs/2021ApJ...917....2X/abstract}
{Xu}, K., {Li}, Q.-C., {Yang}, Y.-P., {et~al.} 2021, \apj, 917, 2, \dodoi{10.3847/1538-4357/ac05ba}

\bibitem[{{Zanazzi} \& {Lai}(2020)}]{https://ui.adsabs.harvard.edu/abs/2020arXiv200205752Z}
{Zanazzi}, J.~J., \& {Lai}, D. 2020, \apjl, 892, L15, \dodoi{10.3847/2041-8213/ab7cdd}

\bibitem[{{Zhang}(2020)}]{2020Natur.587...45Z}
{Zhang}, B. 2020, \nat, 587, 45, \dodoi{10.1038/s41586-020-2828-1}

\end{thebibliography}
\appendix
\section{Exposure Analysis and Period Validation}
The periodic analysis of FRB signals detected by CHIME requires careful consideration of potential aliasing effects due to the telescope's constrained daily observing window of approximately 74 minutes. The detected periodicity of $P_0 = 126.6$ days, corresponding to a frequency of $f_0 = 1/P_0 = 0.0079$ day$^{-1}$, could potentially be aliased with frequencies given by $f_n = N f_{\rm sid} \pm f_0$ \citep{https://arxiv.org/abs/2105.11445}, where $N$ represents an integer and $f_{\rm sid} = (0.99727~{\rm day})^{-1}$ denotes the sidereal frequency. For the first-order alias ($N=1$), this yields potential frequencies of $f_1 \approx 0.995$ or $1.011$ day$^{-1}$ (corresponding to periods of approximately one day), while second-order aliasing ($N=2$) would result in frequencies of $f_2 \approx 1.998$ or $2.014$ day$^{-1}$ (periods of approximately half a day). Our independent confirmation using different methods and clear signs of clustered burst activity makes the possible periodicity a valid case and not an artifact of aliasing. 

Our analysis solely depends on the CHIME-FRB reported bursts. \cite{2024ATel16701....1L} monitored the source with the Very Large Array in two 2-hour segments on 2024-07-02 and 2024-07-05, even after the claimed active phase, and found no bursts at the observing frequency of 1-2 GHz. After 350 hours of monitoring, \cite{2024ATel16732....1O} used the Westerbork RT-1 telescope to discover a single burst at 1.3 GHz at the peak activity of FRB 20240209A in October 2024. This may point to very  frequency-specific activity for the FRB 20240209A. We encourage sensitive radio interferometers which are not bound in exposure times, to observe this source both in intensity and polarization in the next predicted activity period to independently confirm periodic activity in the FRB 20240209A.

\begin{figure}[h!]
    \plotone{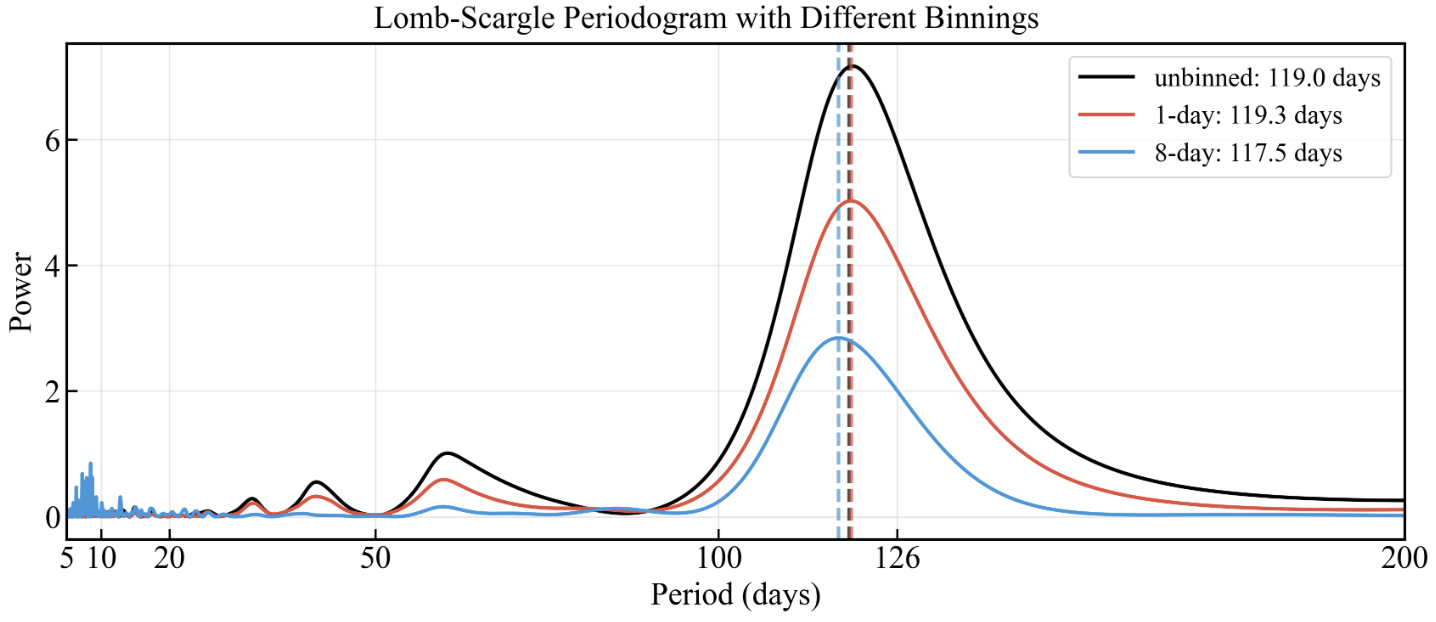}
    \caption{Comparison of Lomb-Scargle Periodogram analysis between unbinned (black), 1 day and (red) 8 days (blue) binning. }
    \label{fig:binnedLSP}
\end{figure}

The FRB shows rapid clustering in the second and third activity epoch, which can lead to faulty estimation of the periodicity and create instabilities in the periodicity detection methods. To properly account for it, we have binned the time of arrivals assuming different averaging factors of 1, and 8 days and ran Lomb-Scargle analysis for all of them. We have allotted the mean time of arrival for each of the sampled bins. From the fig. \ref{fig:binnedLSP} and table \ref{tab:period}, all of the binned methods produce consistent results and hint towards a periodic activity of $\sim 120$ days.
\begin{table*}[h!]
    \centering
    \begin{tabular}{c|c}
    \hline
    Method&Periodicity\\
    \hline \hline
     Autocorrelation Functions  &  $128.5 \pm 6$ days\\
      Epoch Folding  &  $128.8 \pm 14$ days\\
      Lomb-Scargle Periodogram (Unbinned)  &  $119.0 \pm 13$ days\\
      Lomb-Scargle Periodogram (1 Day Binned)  &  $119.3 \pm 14$ days\\
      Lomb-Scargle Periodogram (8 Days Binned)  &  $117.5 \pm 14$ days\\
      \hline
    \end{tabular}
    \caption{Summary table showing all the explored periodicity finding methods and their results}
    \label{tab:period}
\end{table*}

\section{Burst Properties}
The burst properties such as their detection dates, detection DM and SNR are mentioned in the following table as taken from the CHIME repeaters' database.
\begin{table*}[ht]
\centering

\begin{tabular}{cccc}
\hline
Burst ID & Event time & DM (pc cm$^{-3}$) & SNR \\
\hline
1 & 2024-02-09 07:10:14.629273 & $179.542^{+1.617}_{-1.617}$ & 15.986 \\
2 & 2024-02-17 06:36:05.165683 & $178.329^{+0.404}_{-0.404}$ & 9.240 \\
3 & 2024-03-01 02:39:49.031907 & $179.542^{+1.617}_{-1.617}$ & 9.549 \\
4 & 2024-03-09 04:29:37.491817 & $175.903^{+0.404}_{-0.404}$ & 16.778 \\
5 & 2024-06-08 10:23:09.135605 & $175.903^{+0.404}_{-0.404}$ & 9.003 \\
6 & 2024-06-08 10:23:09.143470 & $179.542^{+1.617}_{-1.617}$ & 9.067 \\
7 & 2024-06-12 08:51:06.703933 & $178.733^{+0.809}_{-0.809}$ & 14.422 \\
8 & 2024-06-12 21:33:20.540160 & $179.542^{+1.617}_{-1.617}$ & 10.885 \\
9 & 2024-06-15 07:55:53.038292 & $179.542^{+1.617}_{-1.617}$ & 11.173 \\
10 & 2024-06-16 22:52:19.363123 & $175.094^{+0.404}_{-0.404}$ & 13.696 \\
11 & 2024-06-19 08:57:36.594419 & $175.903^{+0.404}_{-0.404}$ & 10.456 \\
12 & 2024-06-19 09:03:37.417282 & $175.498^{+0.809}_{-0.809}$ & 10.730 \\
13 & 2024-06-19 10:11:00.873141 & $177.116^{+0.809}_{-0.809}$ & 10.305 \\
14 & 2024-06-20 08:55:51.754186 & $177.116^{+0.809}_{-0.809}$ & 8.467 \\
15 & 2024-06-20 09:14:46.669934 & $177.116^{+0.809}_{-0.809}$ & 8.530 \\
16 & 2024-06-21 21:00:56.070438 & $178.733^{+0.809}_{-0.809}$ & 9.019 \\
17 & 2024-06-25 20:20:30.661627 & $175.498^{+0.809}_{-0.809}$ & 12.294 \\
18 & 2024-06-28 22:05:21.265331 & $177.116^{+0.809}_{-0.809}$ & 9.587 \\
19 & 2024-06-29 07:58:35.964083 & $177.116^{+0.809}_{-0.809}$ & 16.166 \\
20 & 2024-06-29 08:32:47.395548 & $175.094^{+0.404}_{-0.404}$ & 20.667 \\
21 & 2024-06-29 08:57:55.017149 & $175.094^{+0.404}_{-0.404}$ & 19.458 \\
22 & 2024-06-29 11:07:02.025088 & $172.263^{+0.809}_{-0.809}$ & 9.700 \\
23 & 2024-06-29 19:16:34.348024 & $177.116^{+0.809}_{-0.809}$ & 16.118 \\
24 & 2024-07-02 08:36:06.603689 & $178.733^{+0.809}_{-0.809}$ & 9.250 \\
25 & 2024-07-16 06:14:15.889723 & $173.881^{+0.809}_{-0.809}$ & 8.604 \\
26 & 2024-10-13 00:51:21.915730 & $175.094^{+0.404}_{-0.404}$ & 10.814 \\
27 & 2024-10-13 15:05:04.347392 & $176.307^{+1.617}_{-1.617}$ & 13.233 \\
28 & 2024-10-15 01:54:43.840624 & $175.094^{+0.404}_{-0.404}$ & 16.650 \\
29 & 2024-10-16 13:03:56.155627 & $177.116^{+0.809}_{-0.809}$ & 11.842 \\
30 & 2024-10-17 13:48:36.063242 & $175.498^{+0.809}_{-0.809}$ & 10.588 \\
\hline
\end{tabular}
\caption{FRB 20240209A Bursts' Properties}
\label{tab:frb_bursts}
\end{table*}

\begin{table*}[ht]
\centering
\begin{tabular}{cccc}
\hline
Burst ID & Event time & DM (pc cm$^{-3}$) & SNR \\
\hline
31 & 2024-10-19 01:33:14.245499 & $175.903^{+0.404}_{-0.404}$ & 9.287 \\
32 & 2024-10-20 01:41:40.400015 & $179.542^{+1.617}_{-1.617}$ & 13.980 \\
33 & 2024-10-21 00:44:11.257702 & $174.285^{+0.404}_{-0.404}$ & 8.588 \\
34 & 2024-10-22 00:40:36.301260 & $177.116^{+0.809}_{-0.809}$ & 14.014 \\
35 & 2024-10-22 12:23:55.770519 & $175.498^{+0.809}_{-0.809}$ & 13.396 \\
36 & 2024-10-24 14:20:08.455547 & $175.903^{+0.404}_{-0.404}$ & 15.214 \\
37 & 2024-10-24 23:17:14.999736 & $175.498^{+0.809}_{-0.809}$ & 14.078 \\
38 & 2024-10-25 12:24:44.518492 & $174.285^{+0.404}_{-0.404}$ & 12.371 \\
39 & 2024-10-26 11:15:13.038387 & $177.116^{+0.809}_{-0.809}$ & 17.424 \\
40 & 2024-10-26 11:29:07.780905 & $175.094^{+0.404}_{-0.404}$ & 9.295 \\
41 & 2024-10-26 12:54:38.974453 & $175.498^{+0.809}_{-0.809}$ & 16.704 \\
42 & 2024-10-27 00:18:52.517918 & $175.498^{+0.809}_{-0.809}$ & 9.461 \\
43 & 2024-10-27 23:47:52.290683 & $175.498^{+0.809}_{-0.809}$ & 8.947 \\
44 & 2024-10-28 01:40:03.574507 & $174.285^{+0.404}_{-0.404}$ & 10.873 \\
45 & 2024-10-28 12:40:33.872658 & $175.498^{+0.809}_{-0.809}$ & 15.379 \\
46 & 2024-10-29 12:55:40.022220 & $175.903^{+0.404}_{-0.404}$ & 15.839 \\
47 & 2024-10-30 11:54:11.757394 & $175.498^{+0.809}_{-0.809}$ & 15.802 \\
48 & 2024-10-30 23:44:04.395468 & $176.307^{+1.617}_{-1.617}$ & 16.868 \\
49 & 2024-10-31 00:13:47.228948 & $174.285^{+0.404}_{-0.404}$ & 9.444 \\
50 & 2024-10-31 01:40:11.420149 & $175.094^{+0.404}_{-0.404}$ & 11.622 \\
51 & 2024-11-01 23:20:35.248916 & $177.116^{+0.809}_{-0.809}$ & 8.638 \\
52 & 2024-11-02 11:36:03.835873 & $175.498^{+0.809}_{-0.809}$ & 9.359 \\
53 & 2024-11-03 01:35:46.483988 & $174.285^{+0.404}_{-0.404}$ & 10.116 \\
54 & 2024-11-10 00:19:39.894546 & $179.542^{+1.617}_{-1.617}$ & 10.872 \\
55 & 2024-11-19 10:53:51.946974 & $175.498^{+0.809}_{-0.809}$ & 10.537 \\
56 & 2024-11-24 00:08:12.294917 & $175.498^{+0.809}_{-0.809}$ & 10.689 \\
57 & 2024-11-26 22:50:30.653373 & $179.542^{+1.617}_{-1.617}$ & 12.537 \\
58 & 2025-01-09 19:47:17.250759 & $176.307^{+1.617}_{-1.617}$ & 12.339 \\
59 & 2025-01-26 19:12:53.510651 & $173.072^{+1.617}_{-1.617}$ & 10.779 \\
60 & 2025-01-28 18:52:21.264112 & $182.777^{+1.617}_{-1.617}$ & 9.378 \\
61 & 2025-02-05 17:37:48.015301 & $175.498^{+0.809}_{-0.809}$ & 8.766 \\
62 & 2025-02-11 07:11:26.067768 & $174.285^{+0.404}_{-0.404}$ & 8.793 \\
63 & 2025-02-19 18:41:39.946511 & $173.477^{+0.404}_{-0.404}$ & 14.402 \\
64 & 2025-02-28 16:34:35.754672 & $175.498^{+0.809}_{-0.809}$ & 12.300 \\
\hline
\end{tabular}
\caption{FRB 20240209A Bursts' Properties}
\label{tab:frb_bursts2}
\end{table*}
\end{document}